# Indiana Law Review

## ESSAY

### BIG TECH'S TIGHTENING GRIP ON INTERNET SPEECH

GREGORY M. DICKINSON*

*Online platforms have completely transformed American social life. They have democratized publication, overthrown old gatekeepers, and given ordinary Americans a fresh voice in politics. But the system is beginning to falter. Control over online speech lies in the hands of a select few—Facebook, Google, and Twitter—who moderate content for the entire nation. It is an impossible task. Americans cannot even agree among themselves what speech should be permitted. And, more importantly, platforms have their own interests at stake: Fringe theories and ugly name-calling drive away users. Moderation is good for business. But platform beautification has consequences for society's unpopular members, whose unsightly voices are silenced in the process. With control over online speech so centralized, online outcasts are left with few avenues for expression.*

*Concentrated private control over important resources is an old problem. Last century, for example, saw the rise of railroads and telephone networks. To ensure access, such entities are treated as common carriers and required to provide equal service to all comers. Perhaps the same should be true for social media. This Essay responds to recent calls from Congress, the Supreme Court, and academia arguing that, like common carriers, online platforms should be required to carry all lawful content. The Essay studies users' and platforms' competing expressive interests, analyzes problematic trends in platforms' censorship practices, and explores the costs of common-carrier regulation before ultimately proposing market expansion and segmentation as an alternate pathway to avoid the economic and social costs of common-carrier regulation.*

---

* Assistant Professor of Law and, by courtesy, Computer Science, St. Thomas University College of Law; Nonresidential Fellow, Stanford Law School, Program in Law, Science & Technology; J.D., Harvard Law School. Genesis M. Perez provided excellent research assistance and valuable suggestions.

INTRODUCTION

These two rights Americans hold dear: the freedom to speak their minds and the freedom not to listen when someone speaking her mind tries to tell them what to do. But over the last two years, social-media platforms have somehow managed to ruin both—curtailing Americans' power to express their views through well-intentioned, but too-powerful content nannying.

## I. SOCIAL MEDIA'S DUELING INTERESTS IN FREE EXPRESSION

To see the first point, consider recent clashes over platforms' content-moderation practices. After months of unproven complaints by Republicans that online platforms disproportionately censor conservative speech,[1] the issue came to a boiling point last year when the nation's leading social media platforms, Facebook, YouTube, and Twitter, all banned then-President Donald Trump.[2] That an online platform would silence a sitting president of the United States is astounding. But, then again, so too are the allegations against him, namely that he instigated the Capitol Riot on January 6, 2021, in which protestors stormed the U.S. Capitol and forced the legislators to flee the House Chamber.[3] Extreme circumstances call for extreme measures. Were the censorship only of former President Trump, or only to prevent violent mobs, or only to remove dangerous misinformation about COVID-19 "cures,"[4] we might chalk it up to unusual times and reputationally sensitive online entities forced to make difficult calls. The platforms are in a tough spot. As Twitter and Facebook are quick to remind us, they face simultaneous pressure from Congress both to promote free expression and to remove more harmful and offensive content.[5] No solution can make everyone happy.

---

1. *See, e.g.*, Emily Cochrane, *Trump Says Social Media is Censoring the Right*, N.Y. TIMES Aug. 18, 2018, at A24; Cat Zakrzewski, *Prominent Conservatives Take Aim at Tech Companies' Biggest Legal Shield,* WASH. POST, Mar. 22, 2019, at A18; Jonathan Easley, *Conservatives Lean Into Warning on 'Wave of Censorship'*, THE HILL (Jan. 27, 2021), https://thehill.com/homenews/media/536000-conservatives-lean-into-warnings-on-wave-of-censorship [perma.cc/48LL-ELM6].

2. Kate Conger & Mike Isaac, *Citing Risk of Violence, Twitter Permanently Suspends Trump*, N.Y. TIMES, Jan. 9, 2021, at A1; Mike Isaac & Kate Conger, *Facebook Bans Trump to Terms End*, N.Y. TIMES, Jan. 8, 2021, at B1; Gerrit De Vynck & Rachel Lerman, *YouTube Suspends Trump from Adding New Content*, WASH. POST, Jan. 14, 2021, at A17. In response, Trump filed lawsuits against Facebook, Twitter, YouTube, and their CEOs, which allege that the suspensions were coerced by the government in violation of the First Amendment. *See* Michael C. Bender & Sarah E. Needleman, *Trump Sues to Restore Social-Media Accounts*, WALL ST. J., July 8, 2021, at A4.

3. *See* Natalie Andrews & Rebecca Ballhaus, *House Looks to Impeach Trump in Days – Resolution Charges President with Inciting Insurrection; Some in GOP Discuss Censure*, WALL ST. J., Jan. 12, 2021, at A1; Andrea Salcedo, *2 Capitol Police Officers Sue Trump for Injuries Suffered in Riot*, WASH. POST, Apr. 1, 2021, at B3; Charlie Savage, *What Legal Trouble Might Trump Be in?*, N.Y. TIMES, Jan. 12, 2021, at A12.

4. *See* Jason Meisner & Dan Hinkel, *Fake Tests, Fantasy Cures: Scammers Ramping Up*, CHI. TRIB., Mar. 26, 2020, at 5 (describing governmental efforts to stop scams involving fake COVID-19 tests and cures that circulated on the internet during the early days of the pandemic).

But online content moderation has crept beyond the exceptional cases of violent mobs and once-in-a-century pandemics. Censorship decisions now even reach Americans' everyday lives and political banter, including discussion of firearms,[6] sex and gender,[7] climate change,[8] and terrorism[9]—topics likely to offend, but which are also central issues on the modern political landscape.[10] Some naturally find such expression offensive. Others do not. Either way, such speech is hardly exceptional. Much of it is the same sort of boorish name-calling and sound-bite-driven argument that one finds on Fox News or MSNBC every evening. Other online speech is a good deal stronger, ridden with profanity, stereotypes, and slurs, but still of the same general sort—not high political theory, but sadly common to American political discourse. In choosing to cultivate the "right" sort of online worlds, by removing, shadow banning, and affixing warning labels to ugly, but commonplace, discussions, Big Tech has made clear that its priority is online beautification, not free speech, and that its aesthetic vision will require the muzzling of much everyday American speech.

---

5. *See* Tony Romm et al., *Lawmakers Clash with Silicon Valley CEOs on Liability,* WASH. POST, Oct. 29, 2020, at A21; Eli Rosenberg, *War on Falsity Spares Infowars*, CHI. TRIB., July 17, 2018, at 3; Cecilia Kang, *Can Tech Companies Silence Hate Speech?*, N.Y. TIMES, Apr. 22, 2019, at B1; Mike Isaac, *Facebook Lets Hate Flourish, Report Finds*, N.Y. TIMES, July 9, 2020 at A1.

6. *See, e.g.*, Keith Wood, *Big Tech Attacks the 2nd Amendment*, GUNS & AMMO (Dec. 9, 2020), https://www.gunsandammo.com/editorial/big-tech-attacks-the-2nd-amendment/386695 [https://perma.cc/T8V5-QJTL] (shadow banning, account suspension, and ad removal of posts and ads by firearms, outdoor-gear, and survival enthusiasts).

7. *See, e.g.*, Georgia Wells, *Writer Sues Twitter Over Ban for Criticizing Transgender People*, WALL ST. J. (Feb. 11, 2019), https://www.wsj.com/articles/writer-sues-twitter-over-ban-for-mocking-transgender-people-11549946725 [https://perma.cc/A2T6-NHYF] (blogger banned from Twitter for violation of platform's harassment rules for tweets stating "How are transwomen not men? What is the difference between a man and a transwoman?" and "Men aren't women").

8. *See, e.g.*, John Stossel, *Social Media Censorship Jeopardizes Future of Free Speech*, AP NEWS (Sept. 27, 2018), https://apnews.com/article/243cd95feb064233b13fc7238e818a71 [perma.cc/W2CD-47RK] (YouTube adding links to further information to videos posted by climate-change skeptics).

9. *See, e.g.*, Vera Eidelman et al., *Time and Again, Social Media Giants Get Content Moderation Wrong*, ACLU (May 17, 2021), https://www.aclu.org/news/free-speech/time-and-again-social-media-giants-get-content-moderation-wrong-silencing-speech-about-al-aqsa-mosque-is-just-the-latest-example [perma.cc/79ZJ-8UJP] (discussing numerous instances in which platforms have improperly labeled speech as promoting terrorism).

10. *See* Eugene Volokh, *Facebook Will Now Ban Criticism of "Concepts, Institutions, Ideas Practices, or Beliefs" When They Risk "Harm Intimidation, or Discrimination" Against Religious, National, or Other Groups,* VOLOKH CONSPIRACY (July 6, 2021, 8:54 PM), https://reason.com/volokh/2021/07/06/facebook-will-now-ban-criticism-of-concepts-institutions-ideas-practices-or-beliefs-when-they-risk-harm-intimidation-or-discrimination-against-religious-national-or-other-groups/ [perma.cc/K6DE-JA2N] (observing that platforms' restrictive speech policies apply to ban election speech, "even when candidates for office are debating these very issues").

Of course, Twitter and Facebook have their own interests at stake. Private entities are generally free to do and say what they like. Why should they be forced to host offensive content or transmit posts by users that they have decided to ban? If Facebook's leaders think that hosting certain types of content will diminish their company's brand or spoil its online community, that is their prerogative, right? Tension between authors and prospective publishers is nothing new. The two often butt heads when a publisher's interest in cultivating a specific brand or readership collides with an author's interest in broadly disseminating her ideas—sometimes even unvarnished, off-topic, or half-baked ones. The traditional arbiter has been the market. Publishers are free to reject content and authors to seek alternative publishers, but both are constrained by market forces. The loudmouthed, sensationalist author may find few takers for her work, and the nitpicky editor may find herself short on prospective authors. Author and publisher are free from censorship and compelled speech, but neither is truly free to do as she pleases.

## II. Too Powerful to Resist

Not so for Big Tech, which has completely upended the long-standing author-publisher stasis. Unlike traditional publishers, the major social media companies—Facebook, Twitter, and Google—truly can do as they please. They enjoy such large user bases and concentrated power over social media that they have felt free to dictate the boundaries of online speech. One obvious danger of concentrated power is the potential for bias. Perhaps observing the leftward leanings of Silicon Valley's elites, many conservatives perceive their posts to be disproportionately flagged for censorship.[11] The greater danger, however, may be the censorship decisions that Big Tech freely admits, even celebrates: platforms' efforts to beautify their social-media landscapes by pruning unsightly content and muzzling its purveyors. However well-intentioned such "tidying up" may be, it risks quashing important segments of public speech, not because they are truly beyond the pale of modern American discourse, but because they are out of step with the aesthetic sensibilities of Big Tech.

Thus lands Big Tech's second blow. Facebook and Twitter surely do have an interest in self expressively curating their online communities. But by doing so, they also dictate how Americans conduct their online lives. Facebook and Twitter demand the First Amendment treatment of traditional content providers like *Cosmopolitan* or *The Economist*, which regularly court, curate, and express themselves through the content they publish. But as market-dominating social-media platforms, they are nothing like that. Big Tech holds such immense power over Americans' data, online social lives, and channels of communication that it speaks with an authority most Americans are powerless to ignore. When *The New*

---

11. *See* Jonathan Easley, *Conservatives Lean into Warning on 'Wave of Censorship'*, THE HILL (Jan. 27, 2021, 6:01 AM), https://thehill.com/homenews/media/536000-conservatives-lean-into-warnings-on-wave-of-censorship [perma.cc/48LL-ELM6]; Romm et al., *supra* note 5, at A21.

*York Times* tells you it has rejected your op-ed piece, you ignore it, hold your head high, and publish elsewhere. But when Facebook and Twitter tell you what to do—that is, on what subjects to speak, from what perspectives, and using which rhetorical devices—no amount of ear plugging will fix the problem. The options are stark: Speak as Facebook and Twitter wish, or be cast from online society.

The predicament defies any simple solution. Contrary to what one hears from Congress,[12] there is no bad guy to punish, just a gigantic tech industry that has become the victim of its own success. Big Tech has given us exactly what we (think we) want—unbridled access to addictive media content and insulated social-media communities that feed us precisely the news we want to hear. But in doing so, it has also acquired immense power to control Americans' personal data, online social lives, and political discourse. Platforms naturally want to control their platforms by eliminating offensive content and cutting off undesirable speakers. It is good for business. But such beautification also silences the important but unsightly voices of society's marginalized and unpopular members.

### III. THE COMMON-CARRIER PROPOSAL

With Congress at a standstill, the issue may soon fall to the judiciary. This past term, in *Biden v. Knight First Amendment Institute at Columbia University*,[13] the Supreme Court took up a case challenging former President Trump's decision to block certain users from responding to his tweets. The plaintiffs argued that Trump's Twitter account was a governmental public forum in which they held a constitutionally protected right to speak.[14] Although the Supreme Court dismissed the challenge as moot given Trump's election loss and return to life as a private citizen,[15] Justice Clarence Thomas took the opportunity to comment on the legal challenges surrounding online speech and social media sites.[16] He observed that although social media platforms "provide avenues for historically unprecedented amounts of speech," never before has "control of so much speech [been] in the hands of a few private parties."[17] Soon, he predicted, the Supreme Court will "have no choice but to address how our legal doctrines apply to highly concentrated, privately owned information infrastructure such as digital platforms."[18]

---

12. *See, e.g.*, JOSH HAWLEY, THE TYRANNY OF BIG TECH 4 (2021) (lamenting that "Big Tech represents today's robber barons, who are draining prosperity and power from the great middle of our society and creating . . . a new oligarchy"); Steve Lohr et al., *Reprimands of Big Tech Cross Aisle*, N.Y. TIMES, July 17, 2019, at B1; Cecilia Kang, *Tech Giants 'Bullied' Us, Rivals Testify*, N.Y. TIMES, Jan. 18, 2020, at B1.

13. 141 S. Ct. 1220 (2021) (mem.).

14. *See id.* at 1221 (Thomas, J., concurring).

15. *See id.* at 1220 (mem.).

16. *Id.* at 1221-27 (Thomas, J., concurring).

17. *Id.* at 1221.

18. *Id.*

On this much, Justice Thomas is surely right: Social media platforms have become so central to individual identity and participation in society that many now view online speech as a right of citizenship.[19] True, Twitter and Facebook are private entities; participation on their platforms is entirely voluntary; and they retain ultimate discretion to ban users or content. But they hold such power that to be banned from social media *feels* like one has been cast from civilized society and barred from public speech altogether—a power held exclusively by the government and historically reserved for only the most extreme provocateurs.

Having diagnosed the problem, Justice Thomas next discusses some of the available legal tools that courts might apply in response. Social media platforms may be new, but concentrated private control over important resources is not a new phenomenon. Private entities are generally free to accept or reject customers and otherwise conduct business as they see fit, but the government has long deemed certain resources so critical to the public good as to warrant limitations on that freedom.[20] Under historical doctrines governing common carriers and public accommodations, entities that control certain important public resources—think railroads, telecommunications companies, shops, hotels and the like—are stripped of discretion to accept or reject customers and required to offer their services to all comers.[21] Such requirements are accepted as necessary to the public good, but burdensome, and applied sparingly, typically where industry incumbents hold significant market power and some economic or social barrier makes a market-driven solution unlikely.[22]

Justice Thomas raises the possibility that a similar principle should be applied to social-media platforms. To reduce bias and free-expression concerns, social-media platforms could be required, like railroads and telephone companies, to support all users equally.[23] Most people, after all, would not want our telephone companies to deny service to select individuals based on their political or religious views or their online comments about those subjects. Perhaps the principle also holds true for social media. But common-carrier and public-accommodations laws are, at best, an imperfect solution to the social-media problem. First, they come with numerous well-understood costs: They interfere with entities' freedom to respond to market demands and reduce consumer welfare;[24] undermine corporate profitability and spending on innovation by increasing compliance and litigation costs;[25] and impose licensing requirements and other barriers to entry that solidify market leaders and benefit incumbents over innovators.[26]

---

19. *See* Packingham v. North Carolina, 137 S. Ct. 1730, 1735-37 (2017) (statute barring sex offenders from social networking sites unconstitutional limit on "legitimate exercise of First Amendment rights," for "the vast democratic forums of the Internet" and "social media in particular" are today "the most important places . . . for the exchange of views").

20. *See Knight First Amendment Inst. at Columbia Univ.*, 141 S. Ct. at 1222-23 (Thomas, J., concurring) (collecting cases).

21. *See id.* at 1223.

22. *See id.* at 1223-24.

23. *See id.* at 1224-26 (citing Adam Candeub, *Bargaining for Free Speech: Common Carriage, Network Neutrality, and Section 230*, 22 YALE J.L. & TECH. 391 (2020)); *see also* Eugene Volokh, *Social Media Platforms as Common Carriers?*, 1 J. FREE SPEECH L. (forthcoming) (exploring the possibility of using common-carrier law to limit social-media platforms' power over political discourse).

Second, it remains to be seen whether social media's ills warrant such an extreme remedy. The costs of common-carrier and public-accommodation laws are sometimes justified—where, for example, a start-up telephone company faces the insurmountable obstacle of laying a competing network of telephone lines.[27] But whether Big Tech's advantages over prospective start-ups pose a comparable hurdle is still an open question. In today's data-driven market, Facebook, Google, Amazon's stockpiles of user data give them a tremendous advantage over potential competitors. They can mine their databases for insights into user behaviors and preferences and use that data to develop and target new products.[28] And the network effect of their enormous existing user base protects them against entrants and gives them a natural market for new products.[29] Why join a new social networking site if few of your friends are there? But their lead is not unassailable. Growing attention to privacy[30] and consumer-privacy legislation will reduce the usefulness of Big Tech's data stores.[31] Indeed, consumers are beginning to pay attention to the privacy implications of their technology usage, and some providers are responding with privacy-friendly initiatives. Even in the shadow of Big Tech, recent years have seen the rise of TikTok,[32] Signal,[33] Parler,[34] and Disney+.[35] And, as WhatsApp's privacy snafus[36] and Quibi's failed launch[37] have shown us, it is possible even for established and well-funded companies to stumble. When they do, nimble upstarts have been ready to take their place.

---

24. *See* Daniel F. Spulber & Christopher S. Yoo, *Mandating Access to Telecom and the Internet: The Hidden Side of* Trinko, 107 COLUM. L. REV. 1822, 1899 (2007) (noting that standardization of network services "decreases welfare by reducing product variety" and that "incompatible networks may simply represent the natural outgrowth of heterogenous consumer preferences.").

25. *See* Mozilla Corp. v. Fed. Commc'ns Comm'n, 940 F.3d 1, 49-50 (D.C. Cir. 2019) (per curiam) (affirming FCC decision to reclassify broadband internet as an information service rather than a common carrier on the ground that more lightly regulating the industry would increase ISPs investment and innovation); *see also* Jay S. Kaplan, *Finding the Middle Ground: A Proposed Solution to the Net Neutrality Debate*, 26 GEO. MASON L. REV. 230, 239 (2018) (collecting studies regarding correlation between increased regulation of ISPs and reduction in capital expenditures and innovation).

26. *See generally* Hilary J. Allen, *Regulatory Sandboxes*, 87 GEO. WASH. L. REV. 579, 587-92 (2019) (reviewing impact of regulatory barriers to entry on start-up financial technology firms); Susan P. Crawford, *The Radio and the Internet*, 23 BERKELEY TECH. L.J. 933 (2008) (discussing difficulty and importance of preserving competition in internet industries dominated by established incumbents).

27. *See* Angela J. Campbell, *Publish or Carriage: Approaches to Analyzing the First Amendment Rights of Telephone Companies*, 70 N.C. L. REV. 1071, 1120-21 (1992) (discussing early history of common carrier law and Congress's decision in 1910 to classify telephone companies as common carriers); Mann-Elkins Act, Pub. L. No. 61-218, § 7, 36 Stat. 539, 544-45 (1910) (superseded by Communications Act of 1934).

28. *See* Tom Wheeler, *Big Tech and Antitrust: Pay Attention to the Math Behind the Curtain*, BROOKINGS (July 31, 2020), https://www.brookings.edu/blog/techtank/2020/07/31/big-tech-and-antitrust-pay-attention-to-the-math-behind-the-curtain/ [https://perma.cc/6NN3-Y57Y].

29. *See* Jason Del Rey, *6 Reasons Smaller Companies Want to Break Up Big Tech*, VOX (Jan. 22, 2020, 11:36 AM), https://www.vox.com/recode/2020/1/22/21070898/big-tech-antitrust-amazon-apple-google-facebook-house-hearing-congress-break-up [https://perma.cc/U3F4-JVUH].

---

30. *See, e.g.*, Reed Albergotti & Elizabeth Dwoskin, *Apple's iPhone Privacy Change Rankles Firms*, WASH. POST, Aug. 30, 2020, at A19.

31. *See* Alan McQuinn & Daniel Castro, *The Costs of Unnecessarily Stringent Federal Data Privacy Law*, INFO. TECH. & INNOVATION FOUND. (Aug. 5, 2019), https://itif.org/publications/2019/08/05/costs-unnecessarily-stringent-federal-data-privacy-law [https://perma.cc/PUX7-KYSM] (estimating the costs to the U.S. economy of implementing a comprehensive federal consumer data privacy law, including the costs to tech companies of reduced access to data and lower ad effectiveness).

32. TikTok is an app and website for users to create and share short-form videos. *See Our Mission*, TIKTOK, https://www.tiktok.com/about [https://perma.cc/U8HE-EPRE] (last visited Sept. 15, 2021); Katie Rogers & Cecilia Kang, *Biden Order Overrides TikTok Ban*, N.Y. TIMES, June 10, 2021, at B1 (discussing Biden decision to reverse Trump administration ban and allow the "wildly popular app," with more than 100 million U.S. users).

33. Signal is a privacy-focused alternative to other messaging tools such as SMS, iMessage, and WhatsApp. *See* https://signal.org [https://perma.cc/ASQ3-74J8] (last visited Sept. 16, 2021); Nicole Nguyen, *Choose an Encrypted Chat App for Privacy*, WALL ST. J., Jan. 14, 2021, at B1 ("users flocked" to Signal following a change in Facebook-owned WhatsApp's privacy policy, making it the top free app download in the Apple and Google app stores).

34. Parler has positioned itself as a more permissive, free-speech-focused micro-blogging alternative to Twitter. *See* PARLER, https://parler.com/main.php [https://perma.cc/HMA4-8LQL] (last visited Sept. 15, 2021); Jack Nicas & Alba Davey, *A Web Haven for Trump Fans Faces the Void*, N.Y. TIMES, Jan. 11, 2021, at A1 (explaining that, before being dropped by Amazon and banned from the Google and Apple app stores, Parler had benefitted from millions of new users joining the platform over concerns about censorship on Facebook and Twitter).

35. Disney+ is an on-demand video-streaming service launched by Disney in 2019, which offers video content from Disney and affiliates Hulu and ESPN. *See Movies, Shows, & Sports*, DISNEY+, https://www.disneyplus.com/welcome/disney-hulu-espn-bundle [https://perma.cc/ZT4H-5YB6] (last visited Sept. 15, 2021); Stephen Battaglio, *It's the Plus Sign's Day in the Sun; Disney+. AppleTV+. Paramount+. How the Symbol is Dominating the Streaming Wars*, L.A. TIMES, Dec. 23, 2020, at E1 (reporting the rapid growth of streaming services during 2020, especially Disney+, which gained 86 million subscribers just one year after its launch).

36. In early 2021, Facebook-owned WhatsApp forced users to accept a new privacy policy as a condition of continued use of the messaging app. Many users switched to alternative messaging apps like Signal, which does not collect information about whom their users connect and communicate with. *See* Mike Isaac, *WhatsApp Pushes Back Privacy Update as Misinformation Spreads and Some Users Flee*, N.Y. TIMES, Jan. 16, 2021, at B5 (explaining that the privacy policies that alarmed users had already been in place since 2016); Brian X. Chen, *What Apple's New Privacy Labels Tell Us*, N.Y. TIMES, Jan. 28, 2021, at B7 (comparing WhatsApp and Signal data collection).

That said, there are signs of danger. Somewhat concerning has been the dominance of Big Tech's major firms. Google and Apple, for example, have wielded their market-controlling Android and iPhone app stores to take an astounding thirty-percent cut of all in-app purchases.[38] Both face ongoing legal challenges from Epic Games, maker of the popular Fortnight game.[39] Epic contends that the iPhone and Android app stores constitute unlawful monopolies that forcibly extract profits from iPhone and Android app developers by prohibiting them from offering customers alternative payment mechanisms.[40] And Amazon's control of the online retail space, both as a platform for third-party sellers and as a seller of Amazon-branded products, has triggered antitrust actions by consumers and regulators who allege that it makes anticompetitive use of seller pricing data.[41] But perhaps this is nothing more than the hurly-burly of a competitive market. Sharp-elbowed or even anticompetitive activity alone would not justify common-carrier regulation. Those practices would need to be shown to interfere with free expression.[42]

Troubling, therefore, has been a series of recent moves by Big Tech that has, intentionally or not, undermined Americans' ability to communicate their ideas. This spring, for example, Facebook barred from its Instagram platform a series of ads by Signal that promoted its privacy-minded alternative to Facebook's WhatsApp and Messenger apps.[43] Facebook also for many months censored

---

37. *See* Benjamin Mullen et al., *Quibi Shuts Down Months After Launch*, WALL ST. J., Oct. 22, 2020, at A1 (detailing the failure of the Quibi video streaming service despite Hollywood backing and experienced leadership).

38. *See* Jack Nicas, *Google to Cut Some App-Store Fees.*, N.Y. TIMES, Mar. 17, 2021, at B3 (describing the Google and Apple stores' 30% commission and their decision in 2021 to lower their fees to 15% for small companies and sales under $1 million per year to "appeas[e] developers and regulators who accuse the companies of abusing their dominance of the smartphone industry").

39. Epic Games Inc. v. Apple Inc., No. 20-cv-5640 (N.D. Cal. Aug. 13, 2020); Epic Games, Inc. v. Google LLC, No. 20-cv-5671 (N.D. Cal. Aug. 13, 2020).

40. *See* Epic Games Inc. v. Apple Inc., No. 20-cv-5640 (N.D. Cal. Aug. 13, 2020); *See* Epic Games, Inc. v. Google LLC, No. 20-cv-5671 (N.D. Cal. Aug. 13, 2020); *see also* Reed Albergotti, *Judge Grills Apple CEO on App Store Stranglehold*, WASH. POST, May 22, 2021, at A1.

41. *See, e.g.*, District of Columbia v. Amazon.com, Inc., No. 2021-CA-1775-B (D.C. Super. Ct. Jun. 1, 2021); *see also* Cat Zakrzewski & Rachel Lerman, *D.C. Attorney General Accuses Amazon of Fixing Prices in Antitrust Complaint*, WASH. POST (May 26, 2021), at A18 (detailing allegations in action brought by D.C. Attorney General alleging price fixing via contractual provisions with third-party sellers and describing a pair of earlier lawsuits filed in 2020 by 48 state attorneys general making similar allegations against Google and Facebook).

42. *See generally* Biden v. Knight First Amendment Inst. at Columbia Univ., 141 S. Ct. 1220, 1223-26 (2021) (Thomas, J., concurring) (arguing that large digital platforms are analogous to historical examples of common-carriers).

43. *See* Katie Canales, *Signal Said Facebook Shut Down Its Advertising Account After the*

discussion of the theory, espoused by former CDC Director Robert Redfield,[44] among others, that COVID-19 escaped from a virology laboratory in Wuhan China that was conducting virus "gain-of-function research."[45] Not until President Biden publicly ordered an intelligence investigation did Facebook allow discussion of the lab-leak theory on its platform.[46] Finally, after the January 6 Capitol Riot, both Google and Apple booted the free-speech-focused microblogging app Parler from their app stores, making the app unavailable for new users.[47] Until then, from the lead-up to the November 2020 presidential

---

*Privacy-Focused Messaging App Tried to Buy Instagram Ads Showing How the Social Media Giant Collects Data*, BUSINESS INSIDER (May 4, 2021, 4:54 PM), https://www.businessinsider.com/facebook-blocked-signal-ads-data-collecting-targeting-2021-5 [https://perma.cc/G7SU-3UV7] (discussing Facebook's decision to block Signal from buying Instagram ads that would have shown users the data that Facebook collects from them and also a prior incident in which Facebook initially blocked a campaign ad by U.S. Sen. Elizabeth Warren that criticized Facebook); Aimee Chanthadavong, *Facebook Bans Signal's Attempt to Run Transparent Instagram Ad Campaign*, ZDNET (May 5, 2021), https://www.zdnet.com/article/facebook-bans-signals-attempt-to-run-transparent-instagram-ad-campaign [https://perma.cc/FY33-UH6H]; *see also The Instagram Ads Facebook Won't Show You*, SIGNAL BLOG (May 4, 2021), https://signal.org/blog/the-instagram-ads-you-will-never-see [https://perma.cc/QL8S-E23Z] (providing examples of the banned ads).

44. *See* Ovetta Wiggins & Erin Cox, *Md. Officials Call for Adviser's Resignation after Comments on Virus's Origin*, WASH. POST, Mar. 27, 2021, at B4 (describing a March 2021 interview in which Redfield shared his view that an inadvertent leak from a laboratory in Wuhan was the most likely origin of the virus and calls by state legislators for Redfield to resign from his position as medical advisor to Maryland's governor).

45. A prominent example is Facebook's decision to designate as "False Information" and block access to a *New York Post* opinion article that speculated COVID-19 may have escaped from a Wuhan lab. *See* Steven W. Mosher, Opinion, *Don't Buy China's Story: The Coronavirus May Have Leaked from a Lab*, N.Y. POST (Feb. 22, 2020), https://nypost.com/2020/02/22/don't-buy-chinas-story-the-coronavirus-may-have-leaked-from-a-lab [https://perma.cc/YR6H-W6EK]; Opinion, *Facebook's COVID Coverup*, N.Y. POST (Jan. 5, 2021, 7:25 PM), https://nypost.com/2021/01/05/facebooks-covid-coverup [https://perma.cc/ZM28-RDBV].

46. *See* Newley Purnell, *Facebook Eases Restrictions on Covid-19 Posts – The Move to Lift the Ban is the Latest in the Company's Efforts to Regulate Content*, WALL ST. J., May 28, 2021, at B4 (noting Facebook's shift in policy to "no longer remove the claim that COVID-19 is man-made or manufactured from our apps" after previously undisclosed U.S. intelligence became public, which indicated that researchers in Wuhan had been hospitalized in November 2019, and after President Biden "ordered a U.S. intelligence inquiry into the" virus's origins).

47. From the lead-up to the November 2020 presidential election through the January 6 Capital Riot, Parler built a reputation as a free-speech alternative to Twitter and became a top download in the Apple App Store and Google Play Store. *See* Jack Nicas & Davey Alba, *A Web Haven for Trump Fans Faces the Void*, N.Y. Times, Jan. 11, 2021, at A1; Gerrit De Vynck et al., *Tech Giants Responded to Events but May Be Hesitant to Act Again Quickly*, WASH. POST, Jan. 13, 2021, at A20 (describing Apple and Google's decisions to remove Parler from their stores as well as actions by Reddit, Shopify, TikTok, and smaller tech companies to distance themselves from Trump by removing access to content).

election through the January 6 Capital Riot, Parler had built a reputation as a free-speech alternative to Twitter and became a top download in the Apple App Store and Google Play Store. Amazon then pulled the plug altogether when it refused to continue providing web-hosting services to the company.[48] Parler has limped along for months, struggling to find a reliable hosting alternative to Amazon,[49] facing lengthy Android[50] and iPhone[51] app-store bans and working to implement content-moderation practices that will appease Google and Apple.

## CONCLUSION

Maybe this is just a blip. Perhaps social media's rapid democratization of publication technology has briefly upset the status quo and we are suffering fleeting limitations on free expression as the market adjusts. But perhaps not. Perhaps a more dangerous trend is in motion. Big Tech's efforts to restrict the dissemination of disfavored information, combined with its growing market dominance—which can prevent dissemination by alternative means—is exactly the combination that could ultimately justify a common-carrier-like must-carry requirement. Social media's availability to all Americans and all viewpoints is that important. But we are not there yet. The problem is still working itself out in the marketplace. Will Parler survive? Is antitrust action merited? Will some new upstart emerge to serve all comers? Can the market loosen Big Tech's grip on speech? I hope so. Common-carrier regulation is strong medicine. It must not be taken hastily.

---

48. *See* Keach Hagey & Tim Higgins, *Apple, Amazon Move to Curb Parler*, WALL. ST. J., Jan. 11, 2021, at A4 (describing Amazon's decision to stop providing "cloud-computing services," which forced Parler's website offline).

49. After it was dropped by Amazon, Parler was turned away by numerous prospective providers. *See* Sarah Polus, *Amazon Suspends Parler from Web Hosting Service*, THE HILL (Jan. 9, 2021), https://thehill.com/homenews/news/533512-amazon-suspends-parler-off-web-hosting-service-reports [https://perma.cc/X85Q-389L] (discussing Amazon's decision to cease providing services to Parler in early January); Brian Fung, *How Parler Is Trying to Get Back Online*, CNN (Jan. 19, 2021), https://www.cnn.com/2021/01/19/tech/parler-web-hosting-amazon/index.html [https://perma.cc/GJS5-YY2Z] (detailing Parler's struggle to find a suitable alternative web-hosting service).

50. Google has continued to ban Parler from its Google Play app store since its removal on January 8, 2021. Google explained that it will not allow the app to return unless Parler "implement[s] robust moderation for egregious content." *See* Jack Nicas, *Far-Right App Under Scrutiny for Failing to Police Itself*, N.Y. TIMES, Jan. 9, 2021, at B4.

51. Parler returned to the Apple App Store on May 17, 2021, four months after its removal. *See* Sarah E. Needleman, *Parler Taps New CEO, Returns to App Store*, WALL ST. J., May 18, 2021, at A4 (reporting Parler's return); Kevin Randall, *Parler Curbs Hate Speech on iPhones*, WASH. POST, May 18, 2021, at A20 (describing changes to the iPhone version of Parler, required by Apple as conditions of its return to the App Store, including the requirement that Parler adopt an artificial intelligence–based content-moderation system to automatically identify and remove posts it labels "hate").